\documentclass[a4paper,prl,11pt,tightenlines,amsmath,floatfix,square,showkeys,superscriptaddress]{revtex4}
\usepackage[latin1]{inputenc}
\usepackage{graphicx}

\begin{document}

\title{The role of computation in complex regulatory networks}

\providecommand{\ICREA}
{ICREA-Complex Systems Lab, Universitat Pompeu Fabra (GRIB), Dr Aiguader 80, 08003 Barcelona, Spain}
\providecommand{\SFI}
{Santa Fe Institute, 1399 Hyde Park Road, Santa Fe NM 87501, USA}
\author{Pau Fernández}
\affiliation{\ICREA}
\author{Ricard V. Solé}
\affiliation{\ICREA}
\affiliation{\SFI}

\begin{abstract}
Biological phenomena differ significantly from physical phenomena. At
the heart of this distinction is the fact that biological entities
have computational abilities and thus they are inherently difficult to
predict. This is the reason why simplified models that provide the
minimal requirements for computation turn out to be very useful to
study networks of many components. In this chapter, we briefly review
the dynamical aspects of models of regulatory networks, discussing
their most salient features, and we also show how these models can give
clues about the way in which networks may organize their capacity to
evolve, by providing simple examples of the implementation of
robustness and modularity.
\end{abstract}


\maketitle 

\section{Introduction}

As has been highlighted by John Hopfield, several key features of
biological systems are not shared by physical systems. The origin of
such difference stems from the relevance that information plays in the
first, which is not shared by the second \cite{Hopfield1994}. Although
living entities follow the laws of physics and chemistry, the fact
that organisms adapt and reproduce introduces an essential ingredient that is
missing in the physical sciences \cite{Hartwell1999}. Due to this
fact, biological structures result from evolutionary pathways and as
such they are contingent \cite{GouldBible}.   

Perhaps the most clear consequence of the role of information is
the observation that biological entities perform {\em computations}:
there is an evolutionary payoff placed on being able to predict the
future. Typically, more complex organisms are better able to cope with
environmental uncertainty because they can compute, i.e. they have
memory or some form of internal plasticity, and they can also make
calculations that determine the appropriate behavior using what they
sense from the outside world.

Computation thus becomes a crucial ingredient when dealing with the
description of biocomplexity and its evolution, because it turns out
to be much more relevant than the underlying physics. Its dynamics is
governed mainly by the transmission, storage and manipulation of
information, a process which is highly nonlinear. This nonlinearity is
well illustrated by the nature of signaling in cells: local events
involving a few molecules can produce a propagating cascade of signals
through the whole system to yield a global response. If we try to make
predictions about the outcomes of these signaling events in general,
we are faced with the inherent unpredictability of computational
systems \cite{Wolfram1985}. It is at this level where computation
becomes central and where idealized models of regulatory networks seem
appropriate enough to capture the essential features at the global
scale.

Cells are probably the most complete example of this traffic of signals
at all levels. They comprise millions of molecules that act coherently 
persisting far from equilibrium by the exchange of matter, energy and
information with the environment. All these molecular processes,
ultimately controlled by genes, take place at different points in space
and time and involve the leading participation of proteins, which act
as the nanomachines that drive cellular dynamics. The cellular network
can be divided into three major self-regulated sub-webs:
\begin{itemize}
\item the {\em genome}, in which genes can affect each other's level
  of expression; 
\item the {\em proteome}, defined by the set of proteins and their
  interactions by physical contact; and 
\item the metabolic network (or the {\em metabolome}), integrated by
  all metabolites and the pathways that link each other. 
\end{itemize}
All these subnetworks are very much intertwined since, for instance,
genes can only affect other genes through special proteins, and some
metabolic pathways, regulated by proteins themselves, may be the very
ones to catalyze the formation of nucleotides, in turn affecting the
process of translation. 

It is not difficult to appreciate the enormous complexity that these
networks can achieve in multicellular organisms, where large
genomes have structural genes associated with at least one regulatory
element and each regulatory element integrates the activity of at
least two other genes. The nature of such networks started to be
understood from the analysis of small prokaryotic regulation
subsystems and the current picture indicates that even the smallest
known webs that shape cellular behavior are indeed very complex
\cite{Davidson2002,Lee2002}.

Luckily, all this extraordinary complexity can be abstracted, at least
at some levels, to simplified models which can help in the study of
the inner-workings of cellular networks. Overall, irrespective of the
particular details, biological systems show a common pattern: some
low-level units produce complex, high-level dynamics coordinating
their activity through local interactions. Thus, despite the many
forms of interaction found at the cellular level, all come down to a
single fact: the state of the elements in the system is a function of
the state of the other elements it interacts with. What models of
network functioning try, therefore, is to understand the basic
properties of general systems composed of units whose interactions are
governed by nonlinear functions. These models, being simplifications,
do not allow to make predictions at the level of the precise state of
particular units. Their average overall behavior, however, can shed
light into the way real cells behave as a system. 

On the other hand, whereas the question of how networks of many components
can achieve global order is very important, it is no less important to
gain an understanding of how such networks could have been assembled
step by step throughout the evolutionary process. It seems sensible to
expect some properties of these networks to directly influence 
their capacity to smoothly integrate the changes that can make them
fitter in the next generation. In this context, technology should
immensely benefit from a deep knowledge of the processes behind
biological evolution, since by design, engineered systems are
not at all susceptible of blind tinkering. It is interesting, therefore,
to explore how the same simplified models used to understand global
dynamics can give hints as to how ``evolvability'' could be put into
practice. 


In summary, in this chapter we will explore the computational
dimension of cellular networks. We will see that biological networks
may be computationally {\em irreducible}, and hence why Boolean units
are appropriate to understand their global properties. We will also
briefly review the most important features of the Kauffman model, and
their implications for computation. Finally, taking advantage of the
Boolean approximation, we will show how important aspects of the
capacity to evolve such as robustness and innovation could be
implemented, through the use of simple, clear examples.  

\section{The evidence for computing networks}

\begin{figure*}
  {\centering \includegraphics[width=12cm]{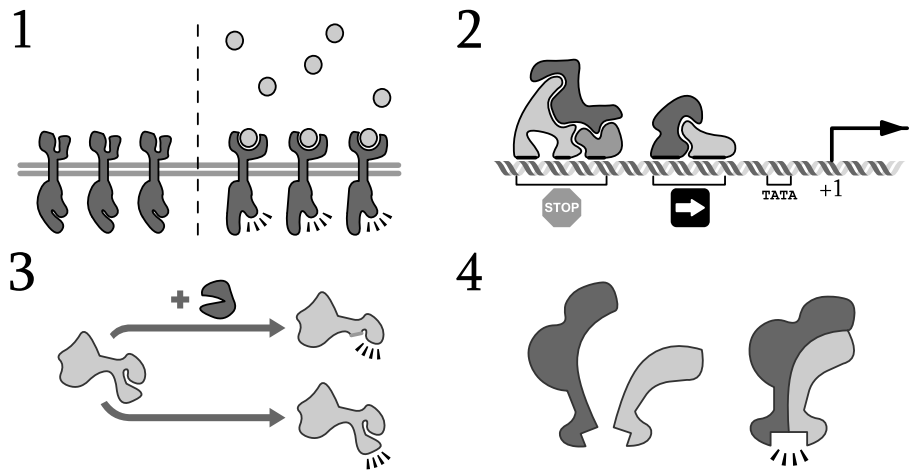}}
  \caption{
    \label{forminter}
    Several ways in which the units of cellular networks interact.
    (1) Signal transduction: a membrane protein becomes active if a
    certain metabolite is present outside the cell. (2) Gene
    regulation: genes are transcribed starting at the $+1$ site and
    are affected positively by an upstream activator sequence and
    negatively by a silencer sequence. Both sequences can be bound by
    protein complexes. (3) Postranslational modification: a given
    protein can be modified after transcription to yield two different
    forms depending on the presence of other proteins. (4)
    Complex formation: the union of two proteins exposes a new active
    site that makes the complex active only then.
  }
\end{figure*}

Molecules, proteins and genes interact with each other in many ways,
and the result of their interactions is the coordinated behavior we
observe. The first step is, therefore, to identify the different
kinds of elements which make up regulatory networks and to describe
their forms of interaction. 

Perhaps the most important units in regulatory networks are
genes, which interact through gene regulation. Genes are translated
into proteins by means of a transcription machinery that is controlled
by multiple mechanisms. Interference with this mechanisms allows
certain molecules to alter the level of expression of specific genes,
as the diagram of figure \ref{forminter}.2 shows. Transcription is
basically initiated at the promoter region, which has usually a
``TATA'' sequence, marking the binding site of TBP (``TATA''-binding
protein). This protein is the first of a series of proteins, known as
general transcription factors, that help to position the RNA
polymerase correctly at the promoter. The most basic regulation,
therefore, involves DNA binding proteins, or regular transcription
factors, that either block the promoter, obstructing transcription or
increase the probability of attachment of the RNA polymerase,
enhancing it. These proteins operate in the vicinity of the promoter
and in a majority of cases form complexes made of many units that
combinatorially bind to DNA.  

In addition to the close binding of transcription factors, other
mechanisms are known that play a significant role in transcription
regulation, including modifications of this basic scheme like
downstream and distal enhancers or totally different mechanisms such
as insulation\cite{Bell2001}, alternative splicing or
post-transcriptional modification\cite{Alberts}. All this mechanisms
affect translation and therefore determine the level of expression of
a certain gene at a given instant, given the concentration of its
multiple regulators.  This level of expression produces a certain
concentration of the protein molecules that are the products of
translation. Actually, different proteins can be produced from the
activation of a certain gene due to post-translational modifications,
giving rise to different regulatory elements in the network. Figure
\ref{forminter}.3 shows an example in which the direct product of a
gene can turn into two different proteins, depending on the presence
of another ``scissor''-like protein that cleaves the initial molecule.

The concentrations of each protein molecule is, however, not only
regulated at the level of transcription. Very many of them have
structures that can be greatly modified in the presence of other
molecules such as metabolites or other proteins. This requires
their separate treatment as regulatory elements, since the different
shapes usually carry out different tasks. Figure \ref{forminter}.1
shows a transmembrane protein which, in the presence of some
metabolite, changes its conformation and becomes active at another
site. These processes are the basis of the functioning of the signaling
network, which comprises membrane receptors, intracellular signaling
proteins and the receivers of the messages, for instance enzymes and
regulatory or cytoskeletal proteins. Many of the components of this
network are proteins that can only be in one of two states, active or
inactive. Other proteins are inactive alone but active while bound to
others in complexes, as shown in figure \ref{forminter}.4, up to very
high levels of complication. As regulatory elements in their own
right, this complexes also qualify as units in the regulatory
network.

\begin{figure*}
  {\centering \includegraphics[width=10.5cm]{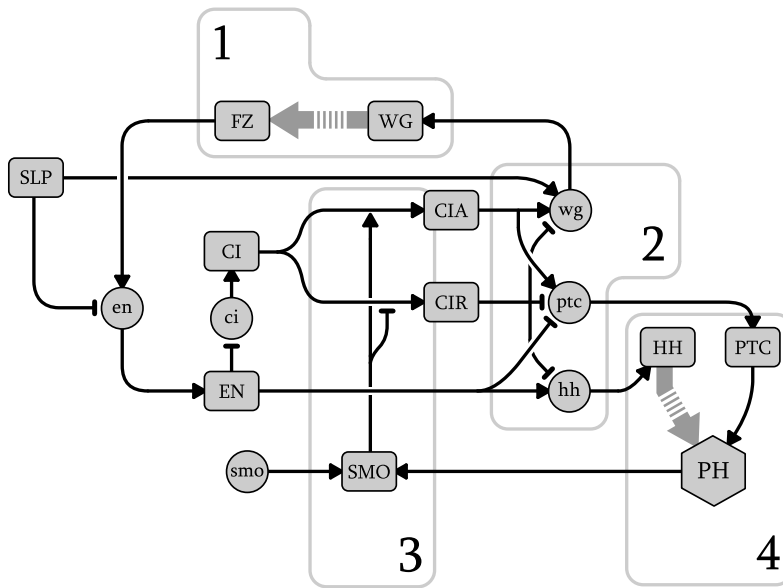}}
  \caption{
    \label{segpol}
    The network of interactions between the segment polarity genes
    (modified from \citep{Albert2003}). Rectangles represent proteins, circles
    genes, and hexagons protein complexes, respectively. Special, thick
    arrows are for transmembrane links. Examples of different types of interaction
    are numbered from 1 to 4. (1) Signaling network: {\em WG} protein binds
    receptor {\em FZ}. (2) Transcription control: {\em wg}, {\em ptc}, and
    {\em hh} transcription is controlled by the activation of other genes. (3)
    Postranslational control: after translation, {\em CI} is transformed into
    {\em CIR} or {\em CIA} depending on the presence of {\em SMO}. (4) Complex
    formation: the complex {\em PH} is formed only when HH and PTC are present.
  }
\end{figure*}

To summarize, many entities in cellular networks can be identified as
the basic units of regulation, mainly distinguished by their unique
roles with respect to interaction with other units. These basic units
are genes, each of the proteins that the genes can produce, each
of the forms of a protein, protein complexes, and all related
metabolites. These units have associated values that either represent
concentrations or levels of activation. This values depend on the
values of the units that affect them due to the mechanisms discussed,
plus some parameters that govern each special form of interaction.

\subsection{Modeling}

To make the description more concrete, is interesting to look at a complete,
real example. In figure \ref{segpol} the circuit of the
segment polarity network of {\em Drosophila melanogaster} is
shown. The genes in this network are expressed throughout the life of
the fly, and its pattern defines and, more importantly, maintains the
borders of the segments since the first stages of development. This is
a network in which all the elements discussed are present, displaying many
forms of interaction, and in particular, the same 4 different
mechanisms depicted in figure \ref{forminter} are highlighted by 4
boxes numbered accordingly. For example, {\em wg} ({\em wingless})
interacts with {\em en} ({\em engrailed}) in the neighboring cells by
secreting a protein, {\em WG}, that binds to a membrane receptor {\em
  FZ} which, when activated, enhances the transcription of {\em
  en}. It is perhaps easier, looking at this diagram, to imagine how
complex the dancing concentrations of genes, proteins or complexes
are, all regulated through their input links and in turn regulating
other elements.

Computer modeling of this network, however, has provided insight into
various questions. A very important result is the fact that this
network seems to be a conserved module. Evidence for
this has been obtained by simulations demonstrating its robustness
against the change of parameters. If the regulatory elements are
modeled using a continuous-valued approach, a set of equations can be
defined governing the rates of change in their populations, levels of
expression, etc. Altogether, the unknown kinetic constants that have to
be specified amounts to 48: half-lives of messenger RNAs and proteins,
binding rates, cooperativity coefficients, etc. Surprisingly, from a
huge number of possible combinations of parameters, many of them have
actually a stable pattern that corresponds to the known pattern of
activity of the genes, thus suggesting that the module is very robust
\cite{vonDassow2000}. In fact, other work has come to similar 
conclusions with respect to other mechanisms, such as adaptive
responses in bacterial chemotaxis \cite{Barkai1997}. It seems that the
topology of the network plays, in some cases, a more important role
than the exact mechanisms at each node.  

This is precisely the thesis of another work \cite{Albert2003}: ``our
purpose here is to demonstrate that in one well-characterized system,
knowledge of the interactions together with their signatures, by which
we mean whether an interaction is activating or inhibiting, is enough
to reproduce the main characteristics of the network dynamics''. The
Boolean network presented, albeit a simplification, seems to
capture the essential features because it matches the patterns of
activity not only of the wild-type embryo, but of some known mutants,
and it also points to other possible effects of mutations that have
not been observed. This is done through an exhaustive analytical
treatment of the resulting equations, as well as simulations, that in
addition reveal the important roles of some of the genes involved 
\cite{Albert2003}. In addition to this, other work approaches gene
networks within the context of the evolution of development
\cite{Sole2003_IJDB}. It is true that the network may not be so
crucial in other cases, but the results nevertheless suggest the
importance of aggregated behavior.   

In brief, the modeling of regulatory networks involves different
methods that give answers to different questions \cite{Hasty2001}, but
ultimately, this methods also illustrate that there is a deep, common
pattern: simulation by computer seems to be the key to the
solutions. As Venter puts it: ``If we hope to understand biology,
instead of looking at one little protein at a time, which is not how
biology works, we will need to understand the integration of thousands
of proteins in a dynamically changing environment. A computer will be
the biologist's number one tool''\footnote{As quoted in \cite{Butler1999}.}.
A crucial question, then, arises: Why do we need a computer
to be able to study biology at all? Some insights into this question
are in fact given by the theory of computation. 

\subsection{Irreducibility}

One of the most important problems in the theory of computation is the
halting problem \cite{Sipser1997}. It concerns the automatic
verification of software in the following terms: one is given a
computer program $A$ and what the program is supposed to do, and the
task is to design an automatic process that verifies the correctness of
the program. In other words, another program $B$ has to be written
that, given a description of $A$ and the correct outputs, predicts
what are the outputs for each set of inputs, and  just checks that the
answers are correct. As simple as the problem seems, it is unsolvable:
there is no such program $B$. The trap lies in the fact that, to solve
it, a computer has to be ``smarter'' than another computer. Instead of
testing the execution of program $A$ by explicitly following it
through, $B$ has to be able to make some kind of shortcut that enables it
to predict the outcome without having to follow each step, to avoid,
for instance, the fact that $A$ may enter a very long or complicated
loop. Since $B$ cannot exist, there are no such shortcuts to the
long-term dynamics of computers, and their step-by-step evolution must
be followed perforce. This impossibility to predict is called {\em
  irreducibility}, and has been hypothesized to be much more common
that usually acknowledged \cite{Wolfram1985}.

The fact that regulatory networks may be irreducible seems to be a
plausible hypothesis, since computer modeling of regulatory networks
seems to be the only way to deal with their complexity. Apart from
that, there seems to be some awareness of this fact, since some authors
have treated cellular networks with the tools of electronic design
\cite{McAdams1995}, and compared molecules with computational
elements\cite{Bray1995}: ``Putting aside for the moment the question
of whether it is useful or even sensible to view them in this manner,
it is nevertheless true that protein molecules are in principle able
to perform a variety of logical or computational tasks''. An
additional reason may be seen in the fact that multistability (or
bistability) is very often the mechanism behind some genetic circuits
\cite{Hasty2001}, and that this switching behavior is the base for
computational capabilities. As a consequence, the assumption that
computational irreducibility characterizes regulatory networks makes
simplified Boolean models sufficient to understand their relevant
properties, since they have the minimal, essential ingredients. This
is the view that we favor in this work. It is important, nevertheless,
to emphasize two important points.

On the one hand, this kind of modeling consciously neglects the details of the
precise functioning of particular units, not because they are
irrelevant, but because they inherently cannot contribute to the
understanding of the whole. The exact strengths of certain interactions
are indeed very important to some physiology processes
\cite{Sveiczer2000}, because this processes determine important aspects
of cell functioning that need fine tuning. But in general, the details
of the switching behavior of networks of many elements do not seem to be
crucial to the overall patterns of activity, which otherwise would
make the network too sensitive to particular parameters. Furthermore, 
irreducibility makes impossible to gain any understanding whatsoever
of a process which involves big numbers of components: ``Even if an
ideal parameter set was provided (say, by software for automatic
parameter optimization), the numerical solutions churned out by the
computer would be just as inscrutable as the cell itself''
\cite{Tyson2001}.

On the other hand, in our opinion, no thorough understanding of all the
processes in the cell can give hints as to why higher level behavior
does occur. After all, understanding means the ability to explain a
phenomenon, which is equivalent to be able to predict its behavior in
all situations. In the case of the whole network, this seems virtually
impossible. Moreover, even if all cell processes were known in detail,
the resulting cell map would be useful for many purposes, such as
designing very complex and specific drugs, but would otherwise leave
open the question of how such a wonderful organization arose through
the accumulation of small variations. An evolutionary explanation of
the assembly of such a complex structure will surely be aided more by
an understanding at the global level of the general dynamics of
idealized Boolean networks than by a detailed study of all the real,
discovered subnetworks. Essentially, there seems to be two levels of
approach in regulatory networks, either at the level of small modules,
or at the level of the whole system, with exclusive goals and
providing answers to qualitatively different
behavior\cite{Anderson1972}. We will focus on Boolean dynamics as the
main tool for whole-system study.

\section{The Boolean idealization}

\begin{figure}
  {\centering \includegraphics[width=6cm]{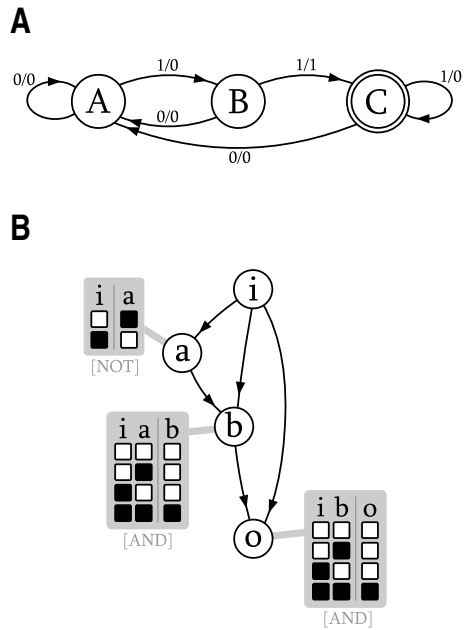}}
  \caption{
    \label{simplest}
    {\bf A}. State diagram of a DFA (Discrete Finite Automaton)
    recognizing the sequence ``011''. The $C$ state represents the
    initial and the final state of the recognition sequence and the
    labels at the links connecting states represent inputs/outputs of
    the automaton. {\bf B}. Minimum Boolean network implementing the task
    defined by the automaton. Each unit in the Boolean network
    computes its next value using a table that tells, for each input
    combination, what should be the output.
  }
 \end{figure}

In order to properly address the role of computation in regulatory
networks, a exactly defined model is required. One possible approach is
to see the networks as devices performing a definite task in
an automatic, orderly manner. Given a set of inputs, such a device
would react by performing a number of predefined operations, and
yielding some output. If the device, either biological or artificial,
has a minimal amount of memory, an appropriate description is provided
by the so called discrete finite automata (DFA), a kind of abstract machines
commonly used in the theory of computation\cite{Sipser1997}. These
automata are also used in the design of logic circuits because they
allow the designer to explicitly state the requirements of a circuit
and they serve as the basis for optimization processes that minimize
various parameters of it, including wiring and the number of memory
units \cite{Hayes1993}.

Figure \ref{simplest}A depicts the state diagram of a DFA, in this
simple case an example of a machine that recognizes the pattern
``011''. This means that given a string of binary digits as input it
will return as output a 1 whenever it detects this pattern, and 0
otherwise. For this purpose, the automaton has three different states
$A$, $B$ and $C$ and at each time step, it is given an input that
makes it jump to another state, and yield some output value. For each
state, two possible transitions are possible, denoted with an arrow in
the diagram. On each arrow two values $a/b$ are drawn, representing
the input value of the transition, $a$, and the resulting value
delivered by the machine, $b$. In our case, the $A$ state represents
the initial phase of the detection, in which the first $0$ is
detected. In fact, all states go to $A$ if a $0$ is
given. Accordingly, $B$ represents the middle phase of the detection
and $C$ the final one, being the initial state as well. 

In figure \ref{simplest}B the simplest network that performs
the task defined by the above automaton is shown. It has one input unit,
$i$, one output unit, $o$, and two internal units, $a$ and $b$. Given
an initial state with all units set to $0$, at each time step, all units
compute their next output as a function of their present inputs,
and switch to the new values at once. For units that have no inputs,
the next value is assumed to be specified. It is not difficult to
trace the values of the units through the detection sequence. First,
upon the reception of a $0$, $a$ switches to $1$, and the other units
remain at $0$. The unit $a$ is then a testimony of a $0$ in the input
at the last time step. At the next time step, provided then that $a$ is
active, $b$ turns to $1$ only if the input was $1$, thus implicitly
detecting a $01$ by means of the temporary memory of $a$. Finally, if
$b$ is $1$ and the input is again $1$, the output turns to $1$,
ending the detection process.  

This network is a simple example of a general class of networks called
Boolean networks, in which inputs perform Boolean functions
\cite{AldanaReview}. The basic ingredients have been used already in the
example: Boolean (i.e. on-off) states for the units, discrete time steps
(synchrony), and general Boolean functions (a different specified
output for each combination of inputs) at each unit. Its introduction
was motivated by the questions raised in the modeling of the gene
regulatory network by Kauffman \cite{Kauffman1969}, although with a
somewhat different perspective. In the last section, we have seen examples of the
modeling of real networks with the aim of understanding particular
parts of the cellular network. Kauffman adopted the complementary
perspective of studying Boolean networks wired at random, with the
hope of finding properties that would apply to the system in its
entirety \cite{Kauffman1993}. As we will see, he mostly succeeded.

Currently known as the Kauffman model, a system composed of $N$ genes
$g_i$ interacting through Boolean functions $f_i$, with discrete
time steps, has a dynamics defined by the following equation:
\begin{equation}
\label{boolnet_dynamics}
g_i^{t+1} = f_i(g_{j_1}^{t},\ldots,g_{j_K}^{t}).
\end{equation}
To fully specify the network, the $K$ inputs of each node are chosen at
random among the $N$ units of the system, and the functions are chosen
so that the outputs have a $1$ with probability $p$ and a $0$ with
probability $1-p$, with no special units as inputs or outputs. Since
it is specified at random, the network only has two parameters of
interest: $K$, which defines the average connectivity between nodes; and $p$, which
actually tunes the susceptibility of the function to changes in the
input values: the closer $p$ to 0.5, the easier it is that $f_i$
changes if input $k$ is reversed.

\begin{figure*}
  {\centering \includegraphics[width=14cm]{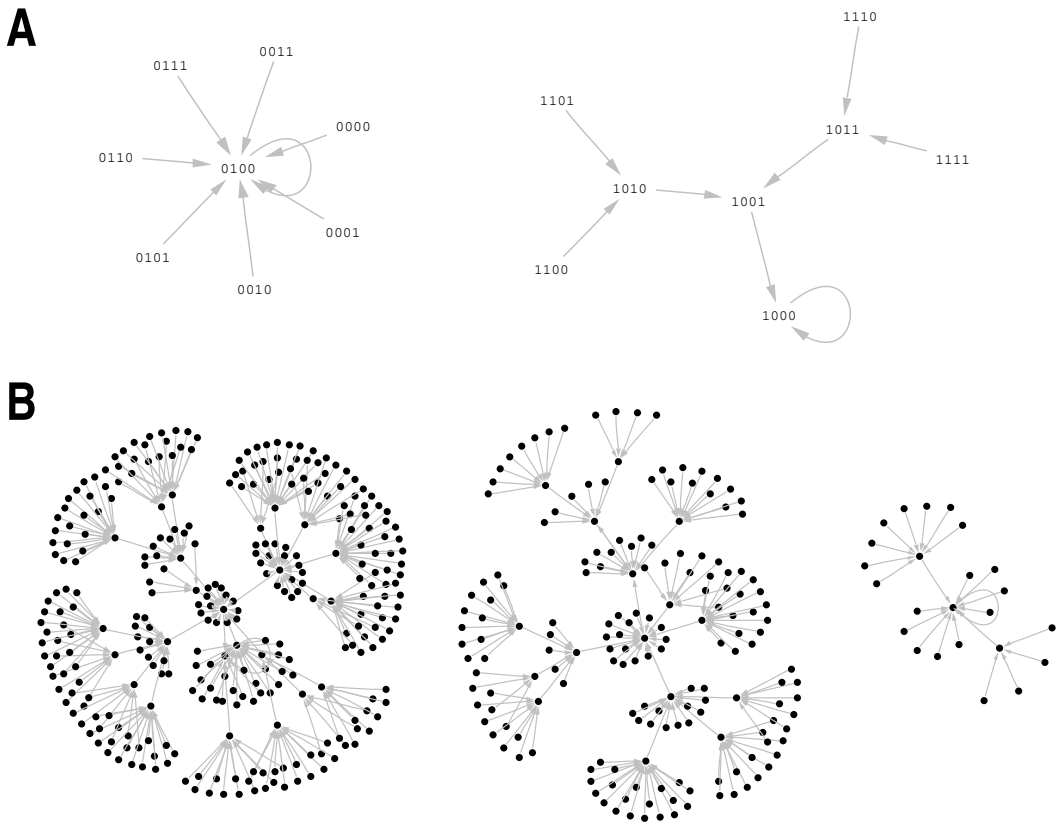}}
  \caption{
    \label{all_basins}
    {\bf A}. Basins of attraction of the circuit shown in figure
    \ref{simplest}. The nodes in this graph are represented as the
    corresponding states of the Boolean network. {\bf B}. Examples of the
    basins of attraction of a randomly generated network with 9 units
    and average connectivity $\langle K\rangle = 2$.
  }
 \end{figure*}

The global dynamics of Kauffman networks can be made clearer
making the following observation. As already mentioned, at each time
step, all nodes are updated synchronously using equation
\ref{boolnet_dynamics} from the values of their inputs. Therefore, we
can treat the whole system as having a global state $S$, given by the
composite state of all the units (or genes), that is,
$$ S \equiv (g_1,g_2,\ldots,g_N). $$ 
This global state $S$ represents a point in the space $\cal{S}$ of all
possible states, and at each time step it jumps to a different point
following a trajectory given by the network configuration, starting at
the chosen initial state. 

Eventually, as time goes on, a state will be reached that has already
been visited before, closing the trajectory into either a loop or a
single state, if this  state maps onto itself. To see the dynamics of
a Boolean network at a glance, it is useful to examine a graph in
which all the possible states of the network are linked to their
successors in the dynamics. In such a chart, disjoint subgraphs
represent different subsets of states that end in the same loop,
called {\em basins of attraction}. Figure \ref{all_basins} depicts
precisely the basin of attraction field of two networks: the example of
figure \ref{simplest} in case A, and a random Boolean network with 9
units and connectivity $K=2$ in case B. All points of the dynamics of
the network are present, followed by their successors, and all possible
trajectories are implicit in them, making the graph a very useful
map. Indeed, software tools exist to draw basins of attraction fields
for any specified network \cite{Wuensche1998}.

Kauffman associated these basins of attraction with the different cell
types specified by the underlying gene network, and made some
arguments regarding the number of cell types (basins of attraction)
present as a function of the network size $N$ and the average
connectivity $K$ \cite{Kauffman1993}. But the most important finding
in relation with our discussion involves the dynamical properties of
the system, and in particular, the propagation of errors. An important
property of Boolean networks is, in fact, that depending on the
connectivity $K$, errors have only three possible fates: either they
die out, propagate to the whole system, or maintain themselves in the
exact border between fading and exploding. This kind
of behavior is a good example of a so called critical phase transition, a
phenomenon well known in statistical physics \cite{Stanley1971}. 

A simple explanation to understand this behavior can be given by means
of a percolation argument \cite{Luque1997}, and is general enough to
include networks with a non-uniform distribution of links of average
$\langle K \rangle$. Consider a given gene $g_i$ in a Kauffman network
of connectivity $\langle K\rangle$, and let us assume that the gene is
externally flipped to the opposite state. The question asked is: how
is this change going to be propagated through the network? Since the
connectivity is $\langle K\rangle$, the change in $g_i$ will arrive, on
average, at the inputs of its $\langle K\rangle$ neighbors. It
remains to be seen with which probability these nodes will propagate
the change, which is the same as asking with what probability a random
Boolean function changes its output when a single input is
changed. Two possible propagation situations can take place, either
the original output was $0$ and shifts to $1$ or the opposite
occurs. Each of these situations has a probability $P=p(1-p)$ (given
the independence between values in the function $f_i$) and two of them
are possible, thus the propagation probability is $P^*=2p(1-p)$. The
average number of changes will be, then,
\begin{equation}
\label{Nch}
N_{ch} = P^* \langle K\rangle = 2p(1-p)\langle K\rangle.
\end{equation}
The three phases of behavior can be understood making the observation
that $N_{ch}$ represents the factor with which errors will
multiply. If $N_{ch}<1$ then changes will tend to disappear, 
at each time step the average number of changes diminishes. This is
the so-called {\em ordered phase}, in which robustness is enough to
cancel errors in the long term. If $N_{ch}>1$ then errors will
multiply and eventually the whole system will be affected by the
avalanche. This is the {\em chaotic phase}, in which the state of
the system in the future is governed by the uniform amplification of
small events. 

At the critical point, that is $N_{ch}=1$, the number of errors does
not have a tendency, so it will be impossible to predict what shall
happen in the long run. In practice, this means that there will be a
mixture of effects: some errors will die out, and some will propagate
to the whole system. Using the equation \ref{Nch}, the critical point
dictates the critical connectivity, 
$$ K_c = \frac{1}{2p(1-p)}, $$ 
which simply leads to $K_c=2$ for the case $p=0.5$, as considered by
Kauffman in its initial formulation. One simple implication of this
formula is the fact that connectivity is rather low, i.e. that the
network is {\em sparse}, an observed property in real networks
\cite{Dorogovtsev2003}. It is also important to note that the
connectivity $\langle K\rangle$ and the probability $p$ alone
determine the global behavior of the system. Although it does 
not make much sense to think that evolution can tune $K$ or $p$
directly, the accumulation of mutations will surely affect them, in
turn affecting its mode of behavior with respect to the phase
transition. 

The importance of this transition lies in its intimate relationship
with computation, and in particular, with the characteristics that
computation requires to systems that implement it. These requirements
have to do with the ability to process information, or in the words of
Langton \cite{Langton1990}: ``First, the physics must support the {\em
  storage} of information, which means that the dynamics must preserve
local state information for arbitrarily long times. Second, the
physics must support the {\em transmission} of information, which means that
the dynamics must provide for the propagation of information in the
form of {\em signals} over arbitrarily long distances. Third, stored and
transmitted information must be able to interact with one another,
resulting in a possible modification of one or the
other''\footnote{Italics from the original.}. In addition, the
issue of irreducibility plays an important role, because systems whose
behavior can be predicted in the long run may not be able to implement
complex tasks.

In the light of these ideas, it does not seem probable that Boolean
networks with computational utility could be in the ordered
phase. Signals do not seem to be able to travel as far as
needed, that is {\em arbitrarily} long distances. Although the
analysis proposed is seen from the viewpoint of errors, a single unit
that serves as input to the system and flips its state can be also
seen as an external signal rather than an error, and then, the
propagation of this error can be regarded as a signaling cascade. If
the signal is unable to reach some parts of the system due to the
network's inherent dynamics, many computations cannot be performed. On
the other hand, computing Boolean networks do not seem to live in the
chaotic phase either. Since regulatory networks are very noisy
\cite{McAdams1999}, any computation that did work in the absence of
noise would be surely disrupted by a single error. The critical phase,
therefore, seems to have the suitable balance: it has the possibility
of communicating any pair of units in the system, and it is not too
sensitive to the values of all of them \cite{Sole1996_FNN}.  

Many authors have drawn attention to the fact that criticality in the
dynamics of Boolean networks or cellular automata have desirable
properties, and in two cases, properties directly related with
computation. The major arguments in favor of criticality are the
following:
\begin{itemize}
\item the capacity of systems at the critical point to exhibit
  arbitrarily large correlation lengths in space and time, supporting
  the basic mechanisms of storage, transmission and modification of
  information \cite{Langton1990}; 
\item the undecidability (or the incapacity to predict without explicitly
  simulating the system) of the properties of systems in the critical
  phase as a basic characteristic of systems capable of computation
  \cite{Wolfram1984}; and, 
\item that emergence of order ``for free'' in networks which are critical
  \cite{Kauffman1993}. 
\end{itemize}
There are also some arguments against this hypothesis. In
\cite{Dhar1995}, it is demonstrated that many cellular automata (a
type of regular Boolean network embedded in space) with computational
capabilities exist in the ordered and chaotic regions defined in
\cite{Wolfram1984}. Their existence is indeed a significant result,
but it does not say anything about the density of automata with
computational capabilities in each phase, which may influence
drastically the probability of reaching them by an  evolutionary
process. In \cite{Aldana2003}, it is argued that Boolean networks with
a scale-free degree distribution may provide, through their uneven
distribution of connectivity, ways of making changes that have a
significant impact on function, but allowing the network to remain
in the ordered phase at the global scale.

Finally, in \cite{Mitchell1993}, an example of simulation
of the evolution of cellular automata is shown that does not select
automata with critical properties, suggesting that the critical phase
does not have a higher density of systems with computational
capabilities. In all cases, it is apparent that evolutionary
properties are a very important ingredient in addition to
dynamics. Overall, however, we are still ignorant about the
applicability of these ideas in real regulatory networks, because
current information includes more data with regard to the presence or
absence of interactions than with their function.

\section{The evolutionary point of view}

To complement current understanding of the dynamics of Boolean
networks, we also want to focus on the functional aspects of network
evolution, again using Boolean networks. Very little is known about
this subject, and yet simple examples can demonstrate the subtle
differences in evolvability between variants of the same circuit.  

\begin{figure*}
  {\centering \includegraphics[width=16cm]{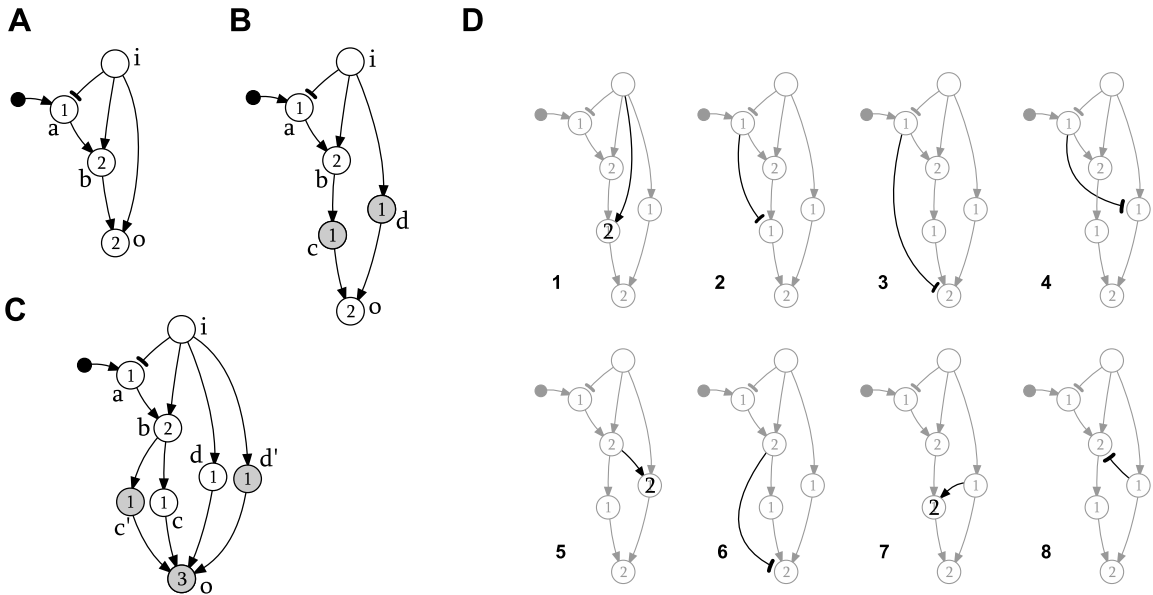}}
  \caption{
    \label{von_neumann_boolnets}
    Several Boolean networks implemented with threshold
    units. Excitatory inputs end in a black arrow, and inhibitory ones
    in a terminating segment. The threshold is specified inside each
    unit, and the units are in gray when they represent changes made
    to another network. {\bf A}. The equivalent of the network of
    figure \ref{simplest}B. {\bf B}. The same functional circuit as in {\bf
    A} with two delay units $c$ and $d$ added. {\bf C}. An example of
    the use of redundancy by the multiplication of lines. {\bf D}. Several
    modifications to {\bf B} that maintain functionality.
  }
\end{figure*}

Figure \ref{von_neumann_boolnets}A shows a Boolean network
implementing the discrete machine shown in \ref{simplest}A. To make
drawings simpler, we have chosen to follow the notation used by
von Neumann\cite{vonNeumann1956}, which eliminates the use of Boolean
tables. Although this notation also implies loosing some richness in
the repertoire of Boolean functions, von Neumann proved its
completeness in the specification of any computational device, and it
is somewhat closer to actual regulation in cells. The new units are also
Boolean and synchronous, but determine their output by comparison to a
threshold $h$. Inputs can either excite or inhibit a given unit, and
the output of the unit at the time $t+1$ will be in the excited state
only if the sum of the number of excitatory inputs minus the number of
inhibitory inputs at $t$ is greater than or equal to $h$. This
threshold $h$ is represented as an integer number inside the circle
that represents the unit. As usual in diagrams of genetic
regulatory circuits, excitation is represented by a black arrowhead
and inhibition by a terminating segment.  

To complete this rather simple scheme we must augment it with the
introduction of a simple unit assumed to be in a permanent excitatory
state, denoted by a smaller black circle, to allow the possibility of
negation. This is in fact what happens with unit $a$ in the diagram of
figure \ref{von_neumann_boolnets}A, to be compared with the circuit of
figure \ref{simplest}B. With respect to the new notation, it is not
difficult to see how the ``AND'' gate behavior of units $b$ and $o$ is
now implemented with units that have two excitatory states and a
threshold $h=2$, since ``AND'' gates are active only if both inputs are
active. Conversely, an ``OR'' unit would be the same as an ``AND'' but
with $h=1$. Finally, before any discussion of the circuit properties
we want to introduce a slight modification to the simplest
implementation. The modification is shown in figure
\ref{von_neumann_boolnets}B, and it just adds a delay of one time unit to
the prior circuit, needing the introduction of units $c$ and $d$ whose
role is simply to retransmit the values at their inputs. The behavior
of the circuit is thus unaltered except for the one time step delay
\footnote{To make the discussion less involved, we have made some
appropriate choices. First, to omit the initial state of the network,
which might give positive output values in the intermediate steps of
the detection. Starting from random values, it is sufficient to
neglect the output in the first steps of the process, and after that
all of it is correct. Second, we have chosen excitation links with
preference, because they make the exposition clearer, although in all
circuits the units can be made to function in reverse with a few
changes.} 

Let us, then, examine this new circuit. One of the first evident
properties is its complete lack of robustness: if the link connecting
$c$ and $o$ happens to fail when transmitting the activation signal at the
final steps of the detection process, $o$ will never activate, and the
overall result will be the failure at recognizing the input sequence
``011''. So it happens with the link from $b$ to $c$, and many others
in the circuit. Boolean logic is unable to cope with the failure of
single components provided that the circuit represents a minimal
implementation, as is the case the circuits we have seen. This
fragility is also displayed by many man-made systems, in which the
failure of individual components is  assumed to be very
infrequent. When a failure finally happens, the system is often not
able to function at all. However, natural systems, and in our case
cells, do have a great deal of robustness, motivated, basically, by
two important sources of distress.  

The first is thermal noise: the same process that makes molecules 
move and wander inside the cytoplasm introduces an inevitable
stochasticity in the effects produced by them, for example at the
level of gene expression \cite{Elowitz2002,Blake2003}. The second, a
byproduct of the first, is mutation: cells inherently accumulate
changes in the genome through time, altering at random the networks
they code for, a source of ``permanent'' noise. Despite the existence
of these two sources of noise, cells behave in a very deterministic
manner, compensating for its presence in some way. Deterministic
responses also may include the explicit exploitation of noise to
generate phenotypic variation, the only exception to its repression. At
the level of molecules, cells have mechanisms to ensure that signals
are received at the appropriate places \cite{McAdams1999}. At the
level of genes, for instance, cells of {\em S. cerevisiae} do not
display signs of a decrease in fitness in a 40 percent of null
mutations to all genes in chromosome V \cite{Smith1996}. The question
is: how can cells achieve this powerful buffering? 

\subsection{Redundancy}

A similar question was probably asked by von Neumann,
albeit in a more abstract manner. He was searching for a logic system
composed of unreliable components which worked in a reliable manner
\cite{vonNeumann1956}. Many engineered systems require, in fact, high
standards of reliability, such as, for instance, computerized bank
accounting. The solution proposed by von Neumann, and still used today
is to put {\em redundancy} into the system, or, stated plainly: to put
many copies of the same thing. The idea is simple: if anyone of the copies
fails, the copies that still work can compensate. In addition, a mechanism is
needed to determine which are the copies that behave correctly. In the
simplest case, the majority rule can be applied, which von
Neumann implemented with his ``majority organ'' \footnote{He called his
  units ``organs''.}. As the name of the rule implies, in the face of
mismatched behavior, the expected correct copies are assumed to be
the most numerous, disregarding the others as wrong. In this way the
failure of the whole system will happen only when, by chance, a number
of copies bigger than half the number of available copies has failed,
an event with a probability that can be made arbitrarily small as the
number of copies grows, compared to the probability of failure of a
single copy. 

Figure \ref{von_neumann_boolnets}C shows our circuit with redundancy
implemented. Nodes $c$ and $d$ have been duplicated to create two
redundant paths, $c'$ and $d'$, one for each signal. \footnote{This duplication
can be interpreted, in fact, as the duplication of genes $c$ and $d$,
a very usual mechanism for the creation of genes in eukaryotes.}. Upon
arrival, $o$ will activate with only 3 of them, allowing the failure
of exactly one. As an example, the probability of failure of the unit
$o$ in this circuit can be calculated if we call $p$ the probability
of failure of its input links. Assuming that failure of links means
not carrying a positive signal, three possible events can occur that
make $o$ fail, which are the failure of 2, 3 or 4 links. Weighting by
the number of combinations in which they can occur, we have the
following equation: 
\begin{equation}
p'=\binom{4}{2}p^2(1-p)^2+\binom{4}{3}p^3(1-p)+\binom{4}{4}p^4.
\end{equation}
With $p=0.1$, the formula yields $p'=0.052$. To obtain a higher gain,
more parallel units could be introduced. It is worth mentioning,
nevertheless, that the redundancy introduced is also useful to absorb
the changes produced by the removal of units, a situation analogous to
the knockout of genes. In the way this circuit is constructed, anyone
of $c$, $c'$, $d$ or $d'$ could be removed with no functional result
whatsoever. In fact, it is clear that many implementations of
networks detecting the pattern ``011'' are possible, each with some
degree of redundancy placed in different points of the
network. Therefore, it seems that the degree to which redundancy is found in
biological systems must be the product of selection, at each
generation adding or removing links that contribute positively or
negatively to the robustness of the organism. 

\subsection{Degeneracy}

\begin{figure}
  {\centering \includegraphics[width=10cm]{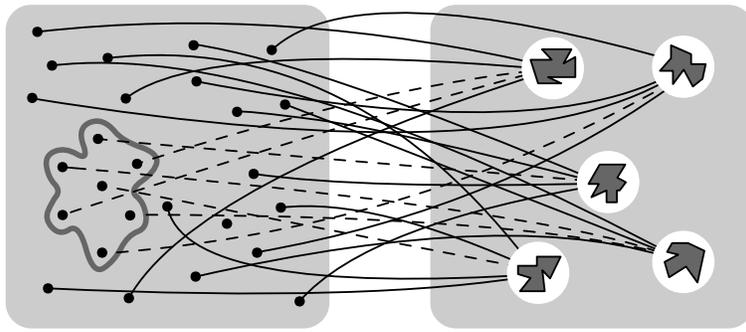}}
  \caption{
    \label{neutral_networks}
    A diagram showing sequences and their associated shapes connected
  by a line. At the left, sequence space, at the right, shape
  space. The neighborhood marked in sequence space has many different
  sequences mapping into all shapes, as the dashed lines reveal.
  }
\end{figure}

But even if simple redundancy seems to suffice for the buffering of
noise or mutation, its utility is of much less relative value than
expected when put in an evolutionary context. Certainly, the
introduction of duplicates protects organisms from mutation and noise,
and studies exist that prove the stability of redundant genes under
some conditions \cite{Nowak1997}. From the perspective of evolution,
however, such a simple form of robustness would make organisms
much less able to innovate. The reason for this difficulty
is that all the copies of subparts that protect redundant systems
probably have to be changed if a change in function is needed, making
the adaptation process very awkward and frustrating. In fact, similar
mechanisms can provide a source of robustness without the drawbacks of
redundancy.

Figure \ref{von_neumann_boolnets}D shows all possible circuits that are 
the same as \ref{von_neumann_boolnets}B, but in which a single link
has been added preserving the global function. In all cases, the new
connections added basically crosscheck the detection sequence of
the units in the simplest circuit. For instance, in the fifth one, unit
$d$ is modified to not only make sure that the last value of the input
is 1, but also its coincidence in time with the activation of $b$,
a detector of ``01'' in the past two values. The path leading from $b$
to $o$ is, in a way, duplicated, because the meanings of $c$ and $d$
overlap to a  certain extent. The other cases involve other parts of
the circuit but result in very similar modifications. These changes, in
fact, can be seen as ``neutral'' mutations. Given the simplest, nude
circuit, different combinations of this single modifications can provide
a great deal of robustness, yet they do so in a different way, taking
advantage of the multiple connections available that do not modify the
behavior of the system. They also seem a more probable source of
robustness, provided that mutation is random in nature.

This mode of robustness has already been defined and has been called
{\em degeneracy}: ``the ability of elements that are structurally
different to perform the same function or yield the same output''
\cite{Edelman2001}. This applies to our system in the sense that
different signaling paths can compute different subparts of the final
pattern without being exact copies. Although first defined in the
context of the nervous system \cite{Tononi1999}, degeneracy seems a
good candidate for the implementation of robustness in biological
systems in general. Redundancy, favored initially due to the existence
of duplication in the genome, was rendered implausible by studies of
duplicated genes showing an immediate and steady divergence of their
sequences, implying that the major source of robustness is to be found
in unrelated genes \cite{Wagner2000}. Again, for the same reasons
mentioned above, the amount of degeneracy can be tuned by evolution to
a suitable degree by making the appropriate changes to the network.

\subsection{Evolvability}

\begin{figure*}
  {\centering \includegraphics[width=14cm]{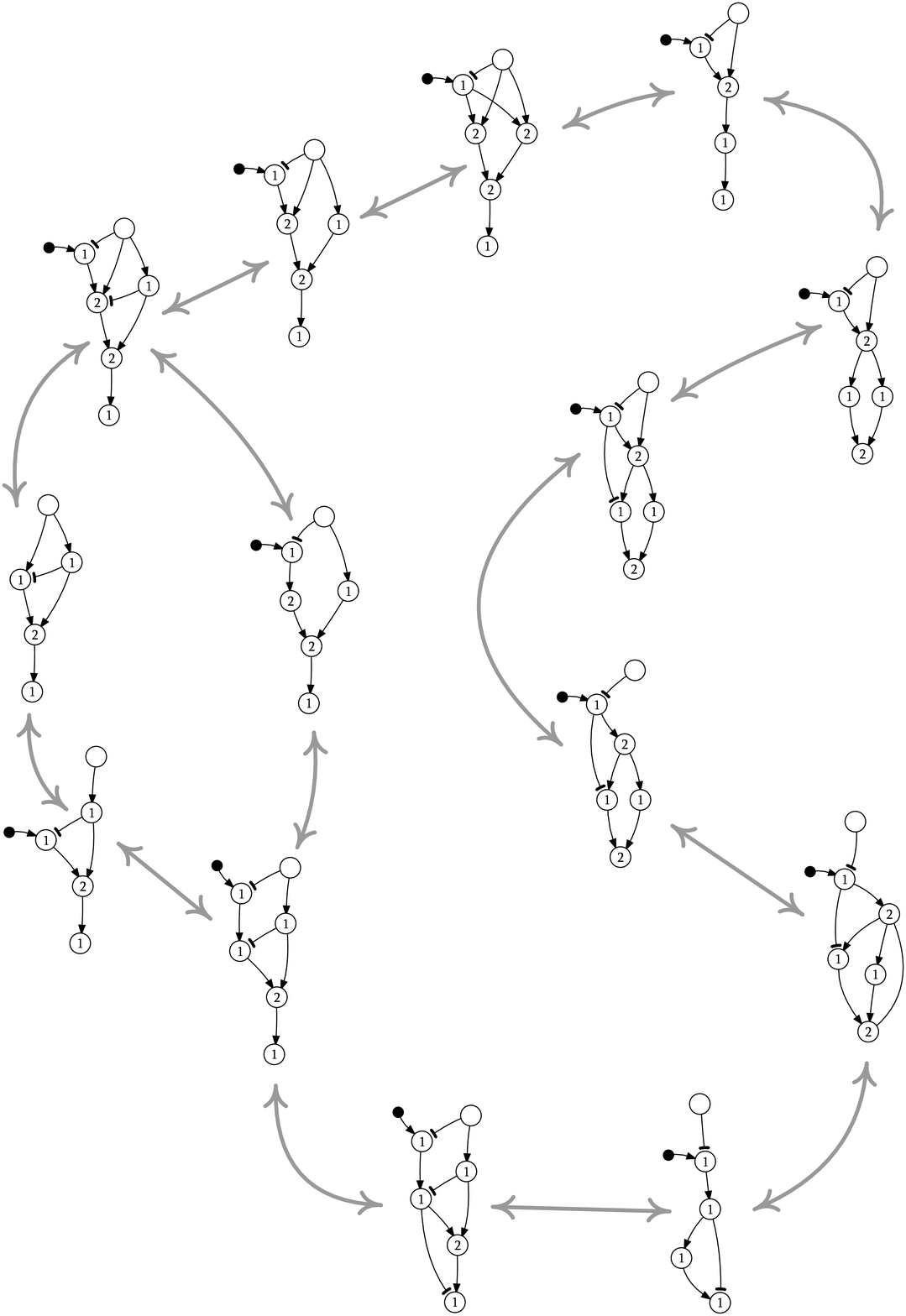}}
  \caption{
    \label{neutral_path}
    An example of a neutral network using simple Boolean circuits. All
  the circuits in the network perform the same task, which is the
  recognition of the ``01'' pattern. On each circuit, the input unit
  is at the top and output is at the bottom. Arrows 
  represent mutations to the circuits, such as duplication or removal of units,
  and addition or removal of links, as well as changes in the
  activation thresholds. 
  }
\end{figure*}

This brings us to the issue of the capacity to evolve, or {\em
evolvability}. Although mentioned only in the context of redundancy,
the inability to innovate is not only related to the duplication of
subparts but also with an excessive display of robustness, even if
implemented using degeneracy. Evolvability has been discussed by many
authors \cite{Kirschner1998,Wagner1996_Evolvability}, and it is
defined as ``the capacity to generate heritable, selectable,
phenotypic variation''. ``This capacity may have two components: (i)
to reduce the potential lethality of mutations and (ii) to reduce the
number of mutations needed to produce phenotypically novel traits'' 
\cite{Kirschner1998}. In relation to our discussion, it is clear that
robustness contributes to the reduction of lethal mutations, but it is
still unclear how to reduce the number of mutations needed to produce
novelty.

Although in a somewhat different context, the study of the evolution
of populations of RNA molecules can provide important insights into this
question \cite{Schuster1996}. In particular, RNA molecules have the
analogs of a genotype and a phenotype in their sequence and folding
shape, respectively. Therefore, a genotype space (or sequence
space) and a phenotype space (or shape space) can be defined. The
studies of the landscapes that appear when linking genotype space with
phenotype space tell us that sequence space is completely
traversed by the so called {\em neutral networks}
\cite{Schuster1994}. These networks comprise all sequences sharing a
common shape that can be accessed by one point mutations, hence the
name. The implications of this fact are more easily understood looking
at figure \ref{neutral_networks}, in which sequence space and shape
space are next to each other. The links that appear between the two
spaces connect sequences with their corresponding shapes. Due to the
existence of neutral networks, all shapes have connections from all
of sequence space. As is immediately apparent, a very small
neighborhood of a given sequence has connections with approximately
all shapes, implying that many shapes are a few mutations away. This
corresponds precisely to the idea of reaching novel traits through a
small number of mutations, our second requirement for evolvability.  

Even if the analogy with RNA has some risks, nothing prevents us in
principle from applying these ideas to Boolean networks. In our
context, sequence space is the analog of our circuit diagram (or genotype),
and shape space is our function space (or phenotype). Neutral
mutations have already been discussed in the context of degeneracy,
where we have seen that many changes to a network do not alter the
network's function. It is therefore a plausible idea that indeed whole
networks of circuits with the same function can be accessed by single
changes in their wiring, enabling a circuit to traverse circuit
space and at the same time undergoing a complete rewiring. This is precisely what
figure \ref{neutral_path} shows. The example circuit has been reduced
to one signaling path that recognizes the subpattern ``01''. Following the
arrows, at each step a single modification is made to the network that
preserves function, including duplications, deletions, and addition or
removal of links. Networks separated by many mutations have very
few common links, and sometimes ``homologous'' links are part of
functionally different signaling pathways.

\subsection{Modularity}

\begin{figure*}
  {\centering \includegraphics[width=13cm]{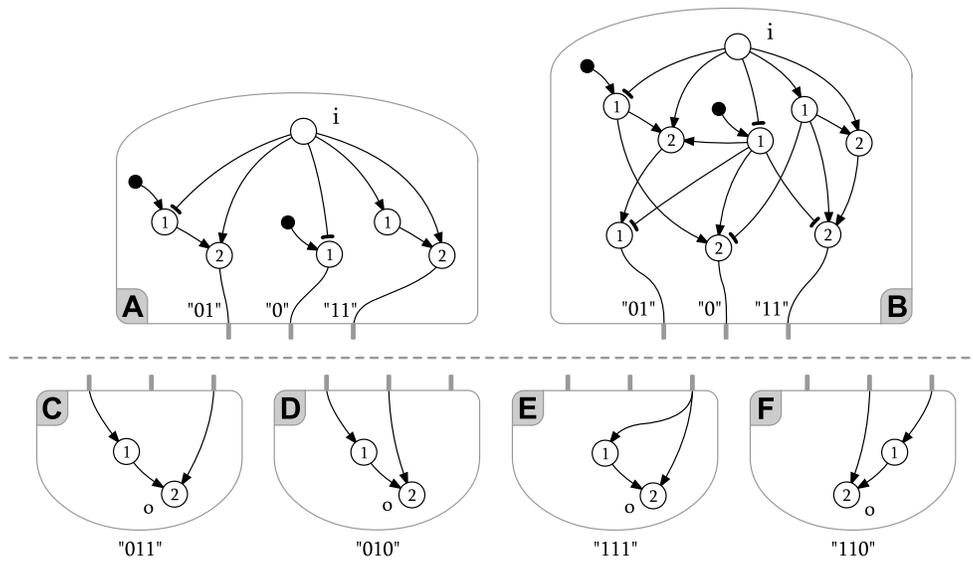}}
  \caption{
    \label{modularity}
    An example of modularity in Boolean networks. {\bf A} Module that
  makes useful ``pre-detections''. {\bf B} The same module with
  added robustness. {\bf C}, {\bf D}, {\bf E} \& {\bf F}. Various ways
  to use the basic module to perform different functions.
  }
\end{figure*}

From another perspective, modularity also seems to contribute to the
successful innovation in organisms. Many examples from
evo-devo show that what makes sense is to study groups of genes in
subnetworks responsible for traits
\cite{vonDassow1999,SoleFYSA,Salazar2001_I} instead of isolated genes,
and from an evolutionary viewpoint, modularity allows the adaptation
of different traits with little or no interference with each other
\cite{Wagner1996_Modularity}. Apart from the separation of
functionally distinct traits, modularity also pervades molecular
biology, with examples such as the recombination of domains in
proteins \cite{Pawson2003}, or the combination of DNA sequences
allowing the cis-regulation of genes \cite{Buchler2003}. In these
cases, what matters most is the recombination of basic modules to form
new structures in a much more rapid fashion, provided that these
modules combinatorially allow any possible higher-level structure to
be built. This gain in speed is widely used in engineering, in which
often new systems are built using the components developed for their
older brothers. Electronics, in particular, is a very good example of
this, with the use of integrated circuits as building blocks that
facilitate the construction of new, more complex circuits.

In figure \ref{modularity}, an example of modularity is shown, again
using the ``011'' detector circuit. As is implied by the shape of 
their boxes, any combination of upper and lower modules can be
plugged to form a different circuit, with the upper modules performing 
basic functions, in this case the detection of particular
subsequences, and the lower modules recombining the outputs of the
upper modules to detect different input patterns. Upper modules
perform the same subfunctions, so as to be compatible with the
interface with lower modules, but differ in the degree of
robustness. This fact illustrates an important point, in relation with
the ideas discussed above, which is the following. As we have already
seen, the degree of robustness can be tuned by an evolutionary
process, giving more or less robustness to selected units in the
network by the addition or removal of ``degenerate'' links. As the
construction of the circuit progresses, a subset of units could be
found to be useful as building blocks for higher-level processing and
thus be made more robust, since the modifications necessary to
generate a complete spectrum of behaviors would not involve these
building blocks but the use of their precalculations in other parts of
the network. Further evolution would be, therefore, speeded up by the
finding of these modules. In this sense, the module $B$ in figure
\ref{modularity} could be one example of that process, resulting in
an increased connectivity within the module.

Modularity in the topological sense is, in fact, measured in terms of
these uneven patterns of connectivity, which produce clusters of nodes
more densely connected. This feature, among others, is what some
methods exploit to detect the modular structure of a
network\cite{Girvan2002,Ihmels2002,Zhou2003}.
A simple enough picture of this kind of modularity, though,
can be obtained by a coefficient $C_i$ which measures the fraction of
neighbors of this node that are neighbors themselves, that is,
$$ C_{i}=\frac{2E_{i}}{k_{i}(k_{i}-1)}. $$
In this formula, $E_{i}$ is the number of edges present between neighbors of
$i$, and $k_i$ the actual number of neighbours, $k_i(k_i-1)$ being the
total number of possible links between neighbors of $i$. The average
of $C_i$, that is, $\langle C\rangle$, describes in general the {\em
 clustering coefficient} of a network. This measure has been observed
to be much higher in real networks than for random graphs in a variety
of fields \cite{Dorogovtsev2003}, and in particular, it has also been
shown to display a scale-free distribution \cite{Ravasz2002}. This
last fact demonstrates that modularity is indeed hierarchical, with
small, strongly connected modules assembling into less cohesive, bigger
modules in the upper level, and so on up to the whole network.

\section{Discussion}

In summary, we are still very much puzzled by the question of how
complex regulatory networks are organized. But we think that the study
of these networks with Boolean models can help understand the properties
of general systems which, on the global scale, behave like real
cells. The reasons for the success of this approximation might be
found in the unsurmountable irreducibility of cellular processes,
which behave in a manner similar to that of a computer. In the case of
particular subnetworks, the Boolean approximation is successful in
studying those mechanisms that are more ``digital'', and do not yield
fine, graded responses. In the case of the whole system, these models
can give important answers to questions regarding global, average
dynamics.

In fact, two important aspects can be readily highlighted about Boolean
networks. On the one hand, their dynamics undergoes a phase transition
that enables us to classify its modes of behavior in three different
zones, depending on a just two global parameters, such as the
connectivity and the unit susceptibility. Looking at the properties of
such modes of functioning, we find more probable that Boolean networks
are in the critical phase, if they are to be capable of
computation. As a consequence, networks must be sparse in
connectivity, a feature which is present in real networks.

On the other hand, simple models of Boolean functions tell us that the
degree of resistance to noise can be varied in a given network, mainly
with the use of degeneracy, which adds neutral connections that
perform parallel processing of the same information. Through a
succession single changes of this kind, a network can be rewired
completely preserving its function at all times. This  resistance to
noise can also be considered as a resistance to mutation, which simply
adds a form of coherent noise to the network. Although good for
robustness, the resistance to change must not be too strong, because
variation is also needed in evolution. Since degeneracy adds
connections and their removal is related to sensitivity, an
equilibrium between the two tendencies seems also to point to the idea
that connectivity in regulatory networks has to be finely tuned to
achieve evolvability. 

In relation to it, modularity might emerge when parts of the network
are found that enable further evolution in a quicker way by
reusing their existing computations. If a suitable combination of
useful modules is found, degeneracy can add protection to them,
increasing connectivity within their subnetworks. This would
implicitly direct the effects of mutations to the connections
governing the combination of modules, which would avoid trying
many worthless mutants. Although these ideas can be presented using
simple examples, much work has to be done to thoroughly quantify them
in models of networks with many units. 







\section{Acknowledgements}
  The authors would like to thank the members of the complex systems 
  research group for useful discussions. This work was supported by
  a grant BFM2001-2154 (RVS) and the Generalitat de Catalunya (PFD,
  2001FI/00732) and The Santa Fe Institute.

\bibliographystyle{plain}
\bibliography{koonin}

\end{document}